\documentclass[preprint,aps,floats,showpacs,amssymb,tightenlines]{revtex4}
\usepackage{epsfig}

\newcommand{\bml}{\begin{mathletters}}
\newcommand{\eml}{\end{mathletters}}

\newcommand{\bea}{\begin{eqnarray}}
\newcommand{\eea}{\end{eqnarray}}
\newcommand{\be}{\begin{equation}}
\newcommand{\ee}{\end{equation}}
\newcommand{\beast}{\begin{eqnarray*}}
\newcommand{\eeast}{\end{eqnarray*}}
\newcommand{\pkt}{\; .}
\newcommand{\kma}{\; ,}

\def\e{{\rm e}}

\begin{document}




\title{Euclidean solutions of Yang-Mills theory coupled to a massive dilaton}
\renewcommand{\thefootnote}{\fnsymbol{footnote}}

\author{Yves Brihaye \footnote{Brihaye@umh.ac.be}}
\affiliation{Dep de Math\'ematiques et Physique Th\'eorique,
Universit\'e de Mons, Place du Parc, 7900 Mons, Belgique}
\author{George Lavrelashvili \footnote{lavrela@itp.unibe.ch }}
\affiliation{Department of Theoretical Physics,\\
A.Razmadze Mathematical Institute,\\
GE-0193 Tbilisi, Georgia}

\date{\today}
\setlength{\footnotesep}{0.5\footnotesep}


\begin{abstract}
The Euclidean version of Yang-Mills theory coupled to a massive dilaton is investigated.
Our analytical and numerical results imply existence of infinite number of branches of
globally regular, spherically symmetric, dyonic type solutions for any values of dilaton mass $m$.
Solutions on different branches are labelled by the number of nodes of gauge field amplitude $W$.
They have finite reduced action and provide new saddle points in the Euclidean path integral.
\end{abstract}

\maketitle

\section{Introduction} \label{intro}%

Motivated mainly by the Bartnik-McKinnon discovery \cite{bm88} of globally regular, static,
spherically symmetric solutions of the Einstein-Yang-Mills (EYM) equations and
similar results obtained in the EYM and YM theory
coupled to a dilaton field \cite{lm93,biz93b,dg93,lm92,biz93a,maison05}
the {\bf Euclidean} version of the EYM and YM theory coupled to a dilaton
were investigated recently \cite{br06}, \cite{bl06}.
While in the four dimensional space-time with Lorentzian signature electric part of
non-abelian $SU(2)$ gauge field should necessarily vanish for asymptotically flat
solutions \cite{eg90,bp92} situation is changed in Euclidean sector: electric field here
plays role similar to the Higgs field of the Lorentzian sector \cite{fg96}
and there are nontrivial dyonic type solutions \cite{br06}, \cite{bl06}.

Dilaton field in these investigations was considered to be massless.
On the other hand there are very strong experimental bounds on dilatonic coupling constant
for the massless dilaton case and it is thought
that at some stage dilaton should obtain potential
possibly with stable or metastable minimum.
Since close to the minimum potential can be well approximated by the dilaton mass term
it is of certain interest to investigate Euclidean solutions in the YM theory coupled
to a {\bf massive} dilaton, which is exactly the aim of present Letter.

The rest of the Letter is organized as follows: in next Section we formulate our model,
derive equations of motion and asymptotic behaviour of finite action solutions.
In Sect. III we discuss some special solutions of our system. In Sect. IV we present
our numerical results. Sect. V contains concluding remarks.

\section{The model} \label{ansatz}%
Starting point of our investigation is the Euclidean version of Yang-Mills theory
coupled to a dilaton field $\phi$, which is defined by the Euclidean action
\be
S_E=\int \left( \frac{1}{2} \partial_\mu \phi \partial^\mu \phi
+ \frac{\e^{2\gamma\phi}}{4 g^2} F_{\mu\nu} F^{\mu\nu}+ V(\phi)
\right)d^4 x \kma
\ee
where $F_{\mu\nu}$ is the non-Abelian gauge field strength, $V(\phi)$ is dilaton field potential
and $\gamma$ and $g$ denote the dilatonic and gauge coupling constants, respectively.

The simplest choice of a dilaton potential would be to have just a mass term:
\be
V(\phi)=\frac{1}{2} m^2\phi^2 \kma
\ee
where $m$ is dilaton mass.

\subsection{The Ansatz and equations}
In this study we will restrict ourselves with the $SU(2)$ gauge group and will be interested in
spherically symmetric solutions.
Following Witten \cite{wit77,fm80} the general spherically symmetric $SU(2)$ Yang-Mills field
can be parameterized as follows:
\be
A_0^a= \frac{x^a}{r} u(r) \kma~\qquad
A_j^a=\epsilon_{jak} x_k \frac{1-W(r)}{r^2}
+\left[\delta_{ja}-\frac{x_j x_a}{r^2}\right]\frac{A_1}{r}
+ \frac{x_j x_a}{r^2} A_2 \kma
\ee
with  $r=\sqrt{x_i^2}$ being the radial coordinate,
$i,j=1,2,3$ are spatial indices and $a,b=1,2,3$ group
indices.

The reduced action $S_{red}$ is defined as follows:
\be
S_E=\int_0^{\tau_{\rm max}} d\tau S_{red} \kma
\ee
with $\tau=x_4$ being Euclidean time. Without loss
of generality we can put $A_1=A_2=0$.
Then the reduced action reads:
\be \label{red_act}
S_{red}= \int dr  {\cal L}_{ymd} \kma
\ee
with
\be \label{lymd}
{\cal L}_{ymd}=\frac{1}{2} r^2 \phi'^2 +
\frac{1}{2} m^2 r^2 \phi^2 + {e^{2\gamma \phi}} {\cal L}_{ym} \kma
\ee
where
\be \label{lym}
{\cal L}_{ym} = \frac{1}{g^2}\bigl(
W'^2+\frac{(1-W^2)^2}{2r^2}+\frac{1}{2} r^2 u'^2 + W^2 u^2 \bigr)
\kma
\ee
and the prime denotes the derivative with respect to $r$.

The equations of motion which follow from the reduced action (\ref{red_act})
read:
\bea \label{eqmW}
W''&=&-2 \gamma \phi' W' -\frac{(1-W^2) W}{r^2} + W u^2 \kma \\ \label{eqmu}
u''&=&-2\frac{u'}{r}-2\gamma \phi' u' +\frac{2}{r^2} W^2 u \kma \\\label{eqmf}
\phi''&=& -\frac{2}{r}\phi'+\frac{2\gamma \e^{2\gamma \phi}}{g^2 r^2}
(W'^2+\frac{(1-W^2)^2}{2 r^2}+\frac{r^2 u'^2}{2}+W^2 u^2 )+m^2 \phi \pkt
\eea

The rescaling
\be \label{rescaling}
\phi \to \frac{\phi}{\gamma}\kma~\qquad r\to \frac{\gamma}{g} r
\kma~\qquad u \to \frac{g}{\gamma} u \kma~\qquad m \to \frac{g}{\gamma} m \kma
\ee
removes the gauge and dilatonic coupling constants $g$ and $\gamma$, respectively,
from the equations of motion. We thus set without loosing generality
$\gamma=g=1$ in the following.

\subsection{Asymptotic behaviour}

Close to $r=0$ we find a 3-parameter family of solutions regular at the origin,
which can be parameterized in terms of $b$, $u_1$ and $\phi_0$:
\bea
\label{zero1}
W(r)&=&1-b r^2 + O(r^4)\kma \\
\label{zero2}
u(r)&=&u_1 r -\frac{u_1}{2} r^2 + O(r^3) \kma  \\
\label{zero3}
\phi(r)&=& \phi_0
+(\frac{1}{6} m^2 \phi_0 + 2 \e^{2\phi_0} \left(b^2+\frac{u_1}{4}\right)) r^2+ O(r^3) \pkt
\eea

The behaviour of the solutions at infinity is more involved.
In leading order we have:
\bea
\label{inf1}
W(r)&=& C \e^{-U r} r^{Q_e} (1+ O(\frac{1}{r})) \kma \\
\label{inf2}
u(r)&=&U-\frac{Q_e}{r} + O(\frac{1}{r^5}) \kma \\
\label{inf3}
\phi(r)&=& -\frac{1+{Q_e}^2}{m^2}\frac{1}{r^4}+ O(\frac{1}{r^6}) \pkt
\eea
So, at infinity solutions are characterised by three free parameters
$U, Q_e$ and $C$.

Note that in the massless case the shift of dilaton field $\phi \to \phi
+ \tilde{\phi}$ for any {\it finite value}
$\tilde{\phi}$ can be compensated by a rescaling of $r$ and $u$ as follows:
\be\label{dilaton_shift}
r \to r \e ^{\tilde{\phi} }\kma~\qquad u\to u \e^{-\tilde{\phi}} \pkt
\ee
This symmetry is obviously broken by dilaton mass term and for finite action solutions
we have to choose normalisation $\phi_{\infty}\equiv \phi (r=\infty)=0$.

\section{Special solutions}

There are few special solutions of our system, which play an important role.

The simplest is the vacuum solution with
\be \label{vacuum}
W(r)=\pm 1 \ \ , \ \  u(r)=0 \ \ , \ \  \phi(r)=0 \ \ .
\ee
Note here that $W(0)=1$ and $W(0)=-1$ are gauge equivalent and in the Eq.~(\ref{zero1})
we have chosen positive sign as in most of the investigations on the subject.

In massless case the equations of motion have elegant solution \cite{biz93a,bl06,fg96},
for which the gauge field amplitudes are simply
\be \label{monopole_g}
W(r)=0 \kma~\qquad  u(r)=U \kma
\ee
with $U$ an arbitrary constant and dilaton field $\phi=\phi_{ab}^{(0)}$
has logarithmic behaviour
\be \label{monopole_d}
\phi_{ab}^{(0)}(r)=-{\rm ln} (1+ \frac{1}{r}) \pkt
\ee
In massive case this abelian solution is modified as follows: gauge amplitudes
$W(r)$ and $u(r)$ are the same as in Eq.~(\ref{monopole_g}) and dilaton field
$\phi=\phi_{ab}^{(m)}$ now satisfy massive equation \cite{fg98}
\be \label{eqmd}
\phi''= -\frac{2}{r}\phi'+\frac{\e^{2\phi}}{r^4}+m^2 \phi \kma
\ee
which unfortunately cannot be integrated analytically.
Solution of this equation $\phi_{ab}^{(m)}$ plays an important role,
while as we discuss below in some limit solution of full system
will be approaching this abelian solution.

In the regime, when dilaton decouples, $\phi \equiv 0$, one finds the BPS monopole solution.
Note that in flat space-time there are no radial excitations of BPS monopole \cite{dm81}
in contrast to case when gravity is taken into account \cite{bfm92,bfm95}.

\section{Numerical results}

We integrated the equations of motions (\ref{eqmW})-(\ref{eqmf}) using the differential equations
solver COLSYS \cite{colsys}. To construct solutions we used the following boundary conditions:
\begin{equation} \label{bczero}
W(0)=1 \ , \ u(0)=0 \ , \  \phi(r)'\vert_{r=0}=0
\end{equation}
at the origin and
\begin{equation} \label{bcinfty}
W(\infty)=0 \ , \ u(\infty) = U \ , \ \phi(\infty)=0
\end{equation}
at infinity.

\begin{figure}[!htb]
\centering
\epsfig{file=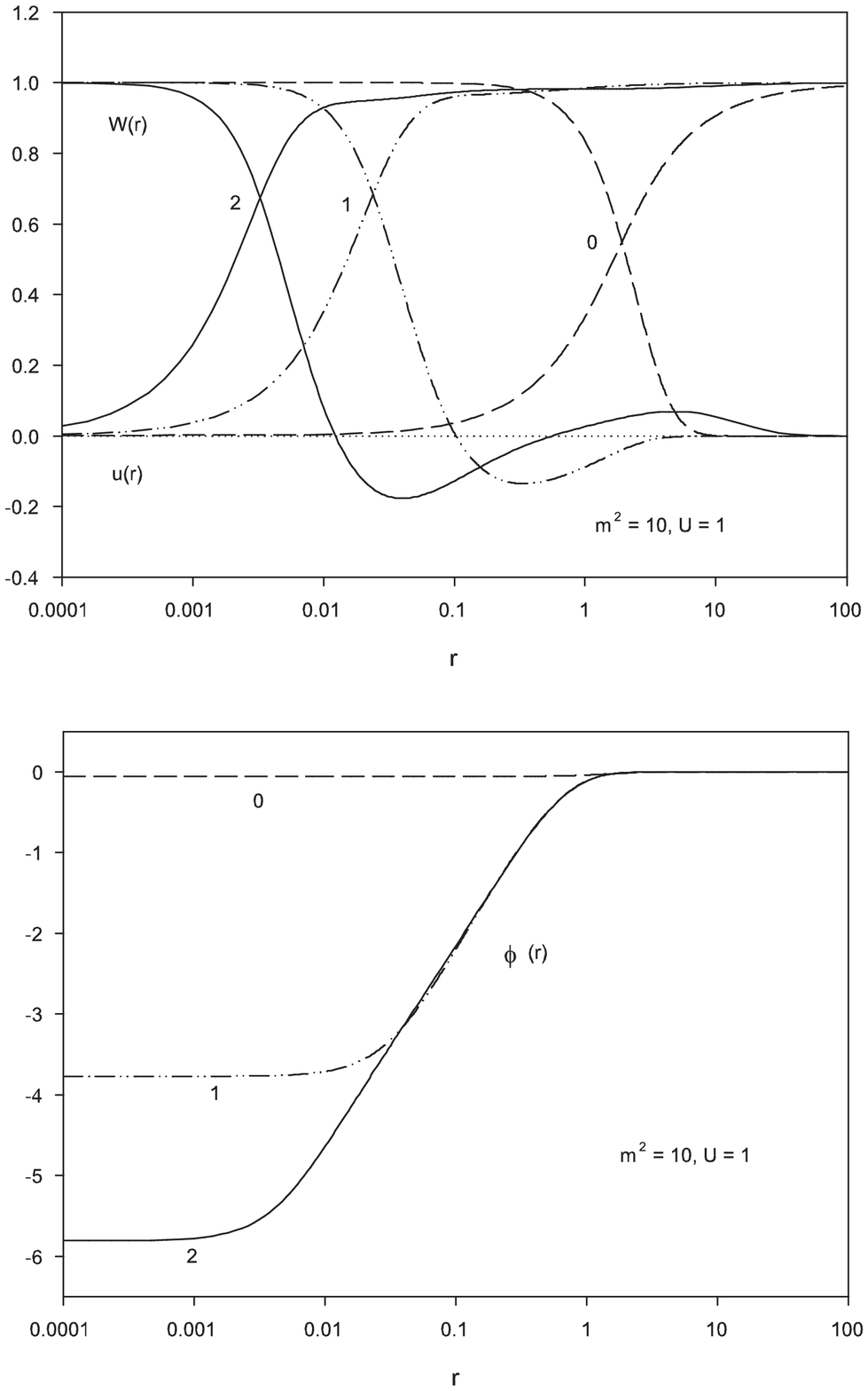,width=0.9\hsize}
\caption{\label{profiles}
Examples of solutions with zero, one and two nodes of the gauge field amplitude $W(r)$
plotted for parameter values $U=1$ and $m^2=10$.}
\end{figure}

{}From numerically found solutions with fixed $U$ and $n$, the number of nodes
of gauge field amplitude $W$, one can extract the values of
$b$, $u_1$ and other parameters which enter in the asymptotic behaviour
Eqs.~(\ref{zero1})-(\ref{zero3}) and Eqs.~(\ref{inf1})-(\ref{inf3}). Examples of solutions with
zero, one and two nodes of the gauge field amplitude $W(r)$ are shown on Fig.~\ref{profiles},
where we plot profiles of $W(r), u(r)$ and $\phi(r)$ for $U=1.0$ and $m^2=10.0$.
Typically dilaton field grows monotonically from some (negative) value $\phi_0$ at $r=0$
to $\phi_{\infty}=0$ at $r=\infty$. Electric component of the YM field, u(r), also grows
monotonically from $u(0)=0$ to some $u(\infty)=U$ and gauge amplitude $W(r)$ starts from its
vacuum value $W(0)=1$ and for the solution with $n$ nodes oscillates $n$ times and then tends
to zero according to Eq.~(\ref{inf1}).

In order to interpret our numerical results, we find it convenient to
analyze the Yang-Mills part of the action density, i.e. the piece
${\cal L}_{ym}$, Eq.~(\ref{lym}), appearing in the full action density Eq.~(\ref{lymd}).
There are two factors which can make $L_{ym}$ large in the neighbourhood of the core of the
soliton: (i) the case $U \gg 1$ , (ii) when many nodes appear, the function $W'$ and/or the ratio
$(1-W^2)/r$ can get large and lead to important contribution to the action density.

The consequence of these features on the pattern of solution is observed already in the massless
case \cite{bl06}. For fixed $n$  and increasing the parameter $U$,
we observe a decreasing of the value $\phi_0$. The factor $exp(2 \phi)$ then get very small
and decreases considerably the contribution to the action of $L_{ym}$ in the region
of the core of the soliton (i.e. for $r\sim 0$).

In the case of a massive dilaton, the situation is more complex because, increasing too much
the value of the parameter $m$ also leads to a large term in the action density and there appears to be a
competition between the Yang-Mills part and the dilaton part.

Note that in problem under consideration we have three different scales:
YMD scale, $l_{ymd}=\frac{\gamma}{g}$, `monopole' scale, $l_{mon}=\frac{1}{U}$ and
scale associated with the dilaton mass, $l_{dilaton}=\frac{1}{m}$.
With the rescaling (\ref{rescaling}) we effectively put $l_{ymd}=1$,
so we still have two scales to vary.
It turns out that the large $m^2$ behaviour of nodeless solutions and solutions with the nodes are
different. So we consider these cases separately.

\subsection{Zero-node solutions}

\begin{figure}[!htb]
\centering
\epsfig{file=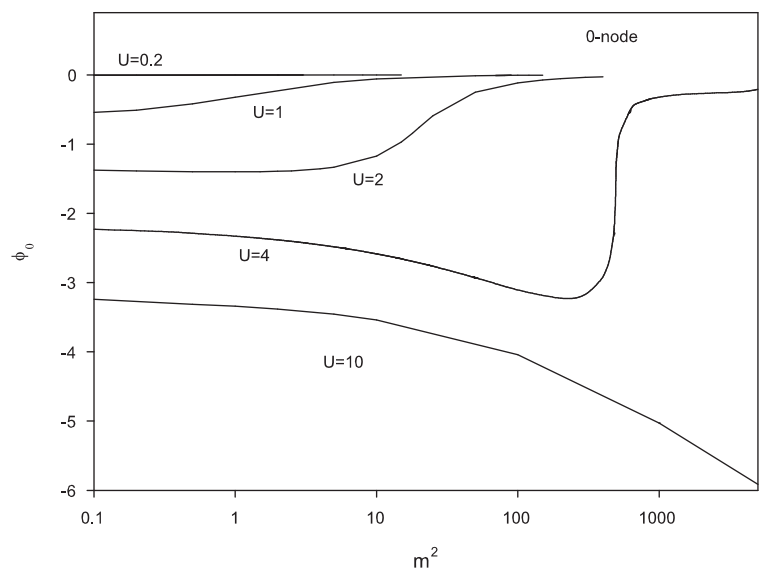,width=0.9\hsize}
\caption{\label{phi0_vs_m2_no_nodes}
Parameter $\phi_0$ is plotted as a function of $m^2$
for different values of $U$ for solutions with zero nodes.}
\end{figure}

The result of the competition between the Yang-Mills field and the massive dilaton is illustrated by
Fig.~\ref{phi0_vs_m2_no_nodes}, where the value of $\phi_0$ is plotted as function of $m^2$ for
several values of $U$ for lowest branch of solutions with no nodes.

For small values of $U$, we see that increasing the mass of the dilaton has the effect to increase
the value $\phi_0$ monotonically. As a consequence, for large enough dilaton masses,
 the function $\phi(r)$ becomes uniformly small
and the corresponding Yang-Mills field converges to the BPS monopole, suitably rescaled,
since $U \neq 1$.  For $U > 1$ the scenario is different, as suggested by
Fig.~\ref{phi0_vs_m2_no_nodes}.
Indeed, increasing $m$, we observe that the value of $\phi_0$ becomes more negative
in the region of the origin, so that the large values of $L_{ym}$ are lowered by the dilaton
factor. While keeping $m$ increasing, we observe that the value $\phi_0$ reaches a minimum
for some $m = m_{cr}$ and then increases to reach zero. We would like to point out that our numerical
analysis becomes rather involved for $m^2 > 1000$. The occurrence of a minimal value for $\phi_0$
can be related to the balance between the Yang-Mills and dilaton pieces of the action density.

\begin{figure}[!htb]
\centering
\epsfig{file=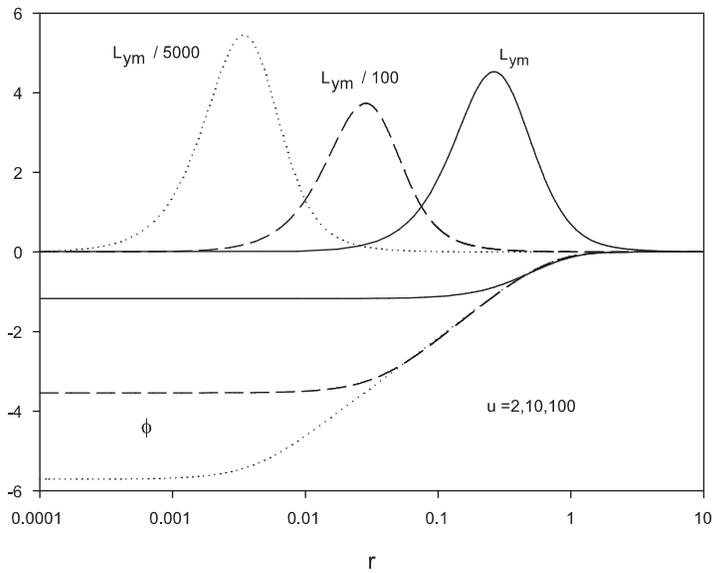,width=0.9\hsize}
\caption{\label{ym_enegy_density}
The Yang-Mills energy density and the dilaton function are plotted as functions of $r$ for
$m^2=10.0$ and $U=2.0,10,100$ respectively by the solid, dashed and dotted lines for
zero nodes solutions.}
\end{figure}

The behaviour of the solutions in the large $U$ limit and with $m>0$ fixed is illustrated
by Fig.~\ref{ym_enegy_density} in the case $n=0$ and $m^2=10$.
Here we reported the Yang-Mills action density
and the dilaton function $\phi(r)$ for three values of $U$, namely $U=2,10,100$. We see
in particular that the function $L_{ym}$ reaches its maximum  closer to the origin while
increasing $U$ defining the radius of the soliton core.
Outside this core the dilaton field approaches the profile $\phi_{ab}^{(m)}$
(which is $m$ dependant) in a larger and larger region of space.
Only inside the core does the dilaton deviates from  $\phi_{ab}^{(m)}$.
Interestingly, our numerical analysis strongly suggest that for large $U$, the relation
\be \label{critical}
\phi_0 = F -  {\rm ln} (U) \ \ , \ \ {\rm with} \ \ F = {\rm const} \ \ \ {\rm for} \ \ \ U \to \infty
\ee
holds, for both cases $m=0$ and $m\neq 0$. In the massless case \cite{bl06} this relation
leads to the existence of critical value $\hat U_{cr}$ when we transform to solutions to
an alternative gauge, say $\hat \phi, \hat U$ where the dilation shift symmetry
Eq.~(\ref{dilaton_shift}) is fixed by demanding $\hat \phi(0)=0$. The critical value
$\hat U_{cr}$ is then determined by $\exp (F)$ where $F$ is the constant above. Our result
strongly suggest that the constant $F$ is independent on $m$; at least it was checked
with a good level of accuracy for $ 0 \leq m^2 \leq 100$.

\begin{figure}[!htb]
\centering
\epsfig{file=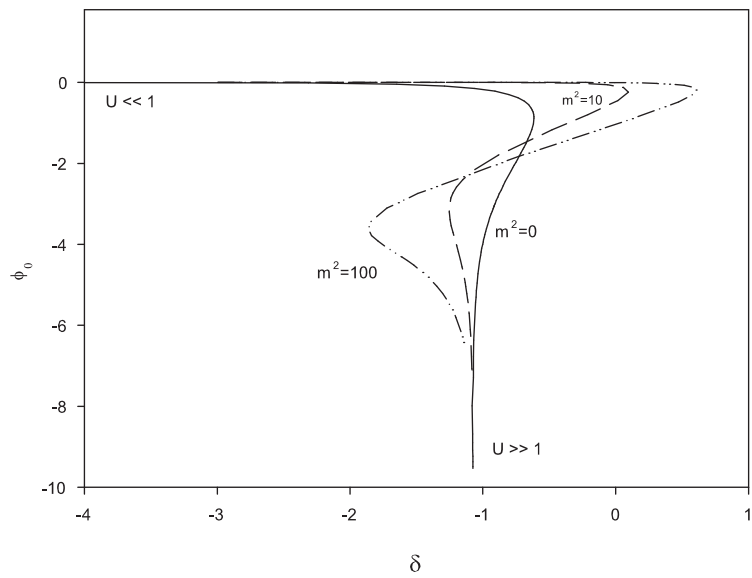,width=0.9\hsize}
\caption{\label{phi0_vs_delta_no_node}
The value $\phi_0$ is plotted as function of ${\rm ln} (U)$
and of the quantity $\delta= \phi_0+{\rm ln} (U)$ for $m^2=0,10,100$ for zero nodes solutions.}
\end{figure}

This remarkable numerical result shows that the critical value are independent on the mass
and simplifies considerably the analysis of the domain of the solutions.
It is illustrated on Fig.~\ref{phi0_vs_delta_no_node}
where the value $\phi_0$ is plotted as a function of the parameter
$\delta \equiv \phi_0+{\rm ln}(U)$ for $m^2 = 0,10,100$. We have for this case $F \approx -1.08$.

\begin{figure}[!htb]
\centering
\epsfig{file=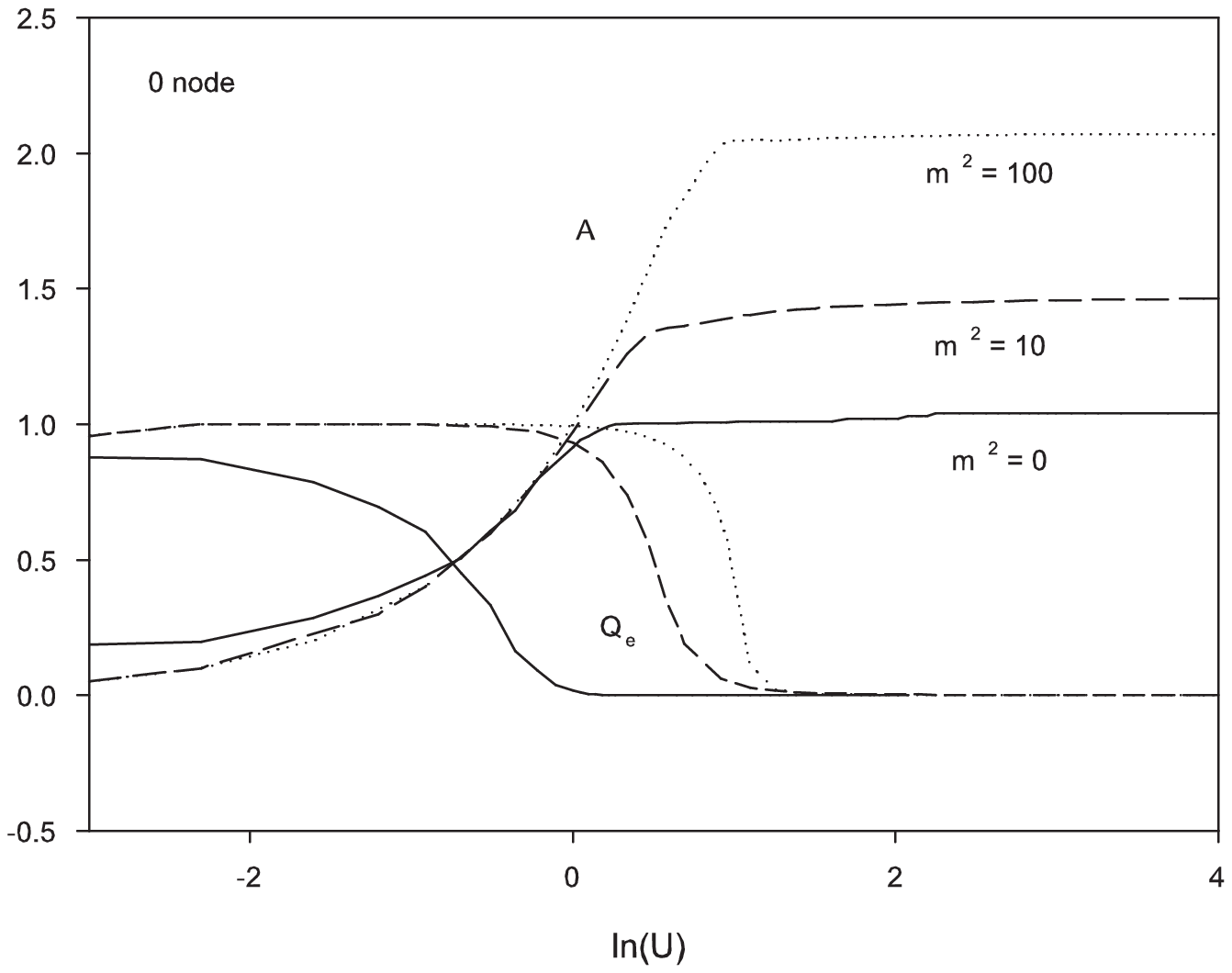,width=0.9\hsize}
\caption{\label{action_no_node}
The value of Euclidean action $A=S_{red}$ and parameter $Q_e$
are plotted as function of ${\rm ln} (U)$ for $m^2=0,10,100$ for zero nodes solutions.}
\end{figure}

The values of Euclidean action $A=S_{red}$ and parameter $Q_e$
are plotted as function of ${\rm ln} (U)$
for $m^2=0,10,100$ for zero nodes solutions on Fig.~(\ref{action_no_node}).

\subsection{Solutions with the nodes}

\begin{figure}[!htb]
\centering
\epsfig{file=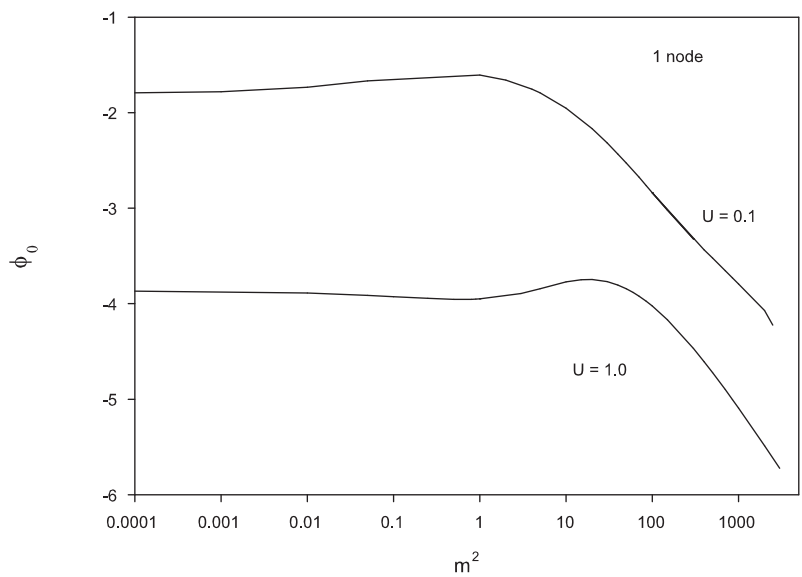,width=0.9\hsize}
\caption{\label{phi0_vs_m2_one_node}
The value of $\phi_0$ is plotted as function of $m^2$ for one node solutions
for $U=0.1, 1.0$.}
\end{figure}

In $m^2 > 0$ case the node solutions constructed in \cite{bl06} get smoothly deformed by the dilaton mass
parameter. Keeping $U$ fixed and increasing the mass, we observe that the value of the dilaton function
at the origin becomes more and more negative. This is due to the fact that, the position of the node,
say $r_0$, of the function $W$, is very close to the origin, so that
the terms $W'$ and $(1-W^2)^2/r^2$ are quite large in this region and favorize negative values of
$\phi(r)$ for $r\in [0,r_0]$. This is illustrated by Fig.~\ref{phi0_vs_m2_one_node}
 where the values $\phi_0$ are reported
as function of $m^2$ for $U=0.1$ and $U=1.0$ (for larger $U$ the pattern is the same).
Note that this behaviour of $\phi_0$ of Fig.~\ref{phi0_vs_m2_one_node} is different
from the zero-node case Fig.~\ref{phi0_vs_m2_no_nodes} and this difference is
explained by absence of radial excitations of BPS monopole in flat space-time \cite{dm81}.

\begin{figure}[!htb]
\centering
\epsfig{file=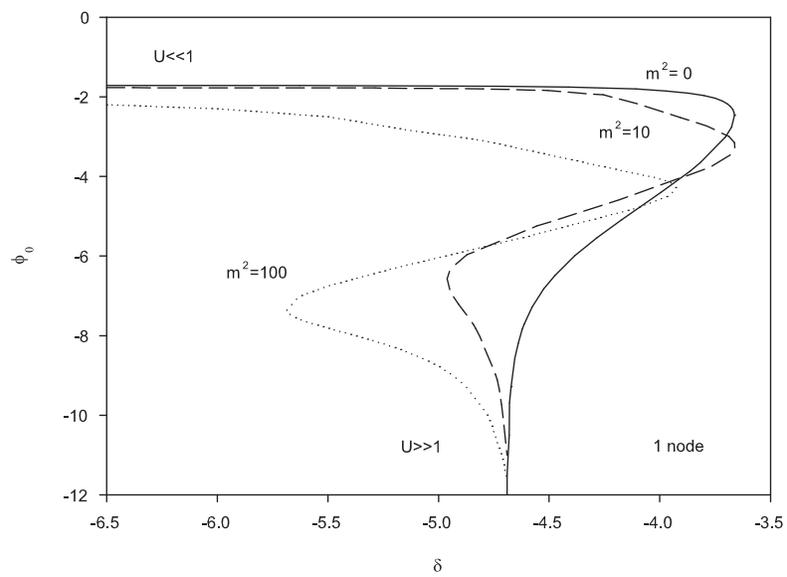,width=0.9\hsize}
\caption{\label{phi0_vs_delta_one_node_}
The value $\phi_0$ is plotted as function of the quantity $\delta= \phi_0+{\rm ln} (U)$
for one node solutions for $m^2=0, 10, 100$.}
\end{figure}

The large $U$-limit of one node solutions was also investigated for several values of $m$
and is illustrated on Fig.~\ref{phi0_vs_delta_one_node_}.
We see in particular on this figure that the property (\ref{critical}) also hold for one-node
solution. Here we find $F \approx -4.70$.

\begin{figure}[!htb]
\centering
\epsfig{file=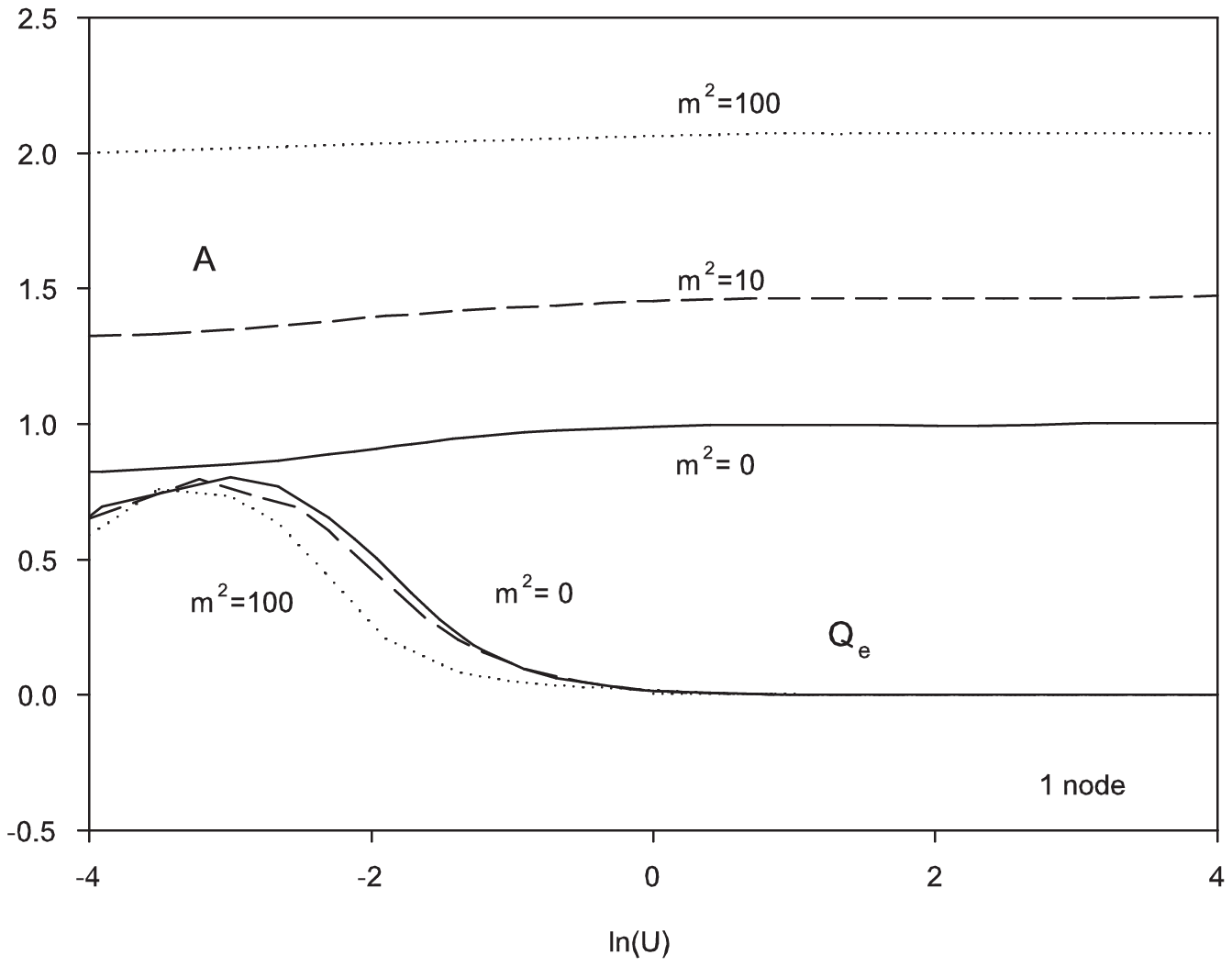,width=0.9\hsize}
\caption{\label{action_one_node}
The value of Euclidean action $A=S_{red}$ and parameter $Q_e$
are plotted as function of ${\rm ln} (U)$
for $m^2=0,10,100$ for one nodes solutions.}
\end{figure}

The values of Euclidean action $A=S_{red}$ and parameter $Q_e$
are plotted as function of ${\rm ln} (U)$
for $m^2=0,10,100$ for one nodes solutions on Fig.~(\ref{action_one_node}).

\section{Concluding remarks} \label{conclusions}

In the present study we investigated $O(3)-$symmetric solutions of
four dimensional Euclidean YMD theory and found branches of dyonic type solutions
for any values of dilaton mass $m$.
They can also be viewed as static (vortex type) solutions  of $4+1$
dimensional theory with Lorentzian signature.
Note that in pure Euclidean YM theory solutions with higher,
$O(4)$ symmetry are well known instantons.
Simple scaling arguments show that there should not be any
finite action $O(4)$ symmetric solutions in combined YM-dilaton theory.

The electric component of YM field in Euclidean theory plays role similar
to the Higgs field and correspondingly present model
is very similar to the static, spherically symmetric
sector of the YM-Higgs model investigated by Forg\'{a}cs and Gy\"{u}r\"{u}si (FG) \cite{fg98}.
The main difference with the FG model is in the
way the dilaton field couples to the the Higgs part of the lagrangian.

It was shown \cite{gl94} that static spherically symmetric solutions of YM dilaton theory
found in \cite{lm92,biz93a,maison05}
have sphaleronic nature similar to their EYM "relatives" \cite{vg91,mw92,gs94},
namely they have unstable mode(s), half-integer topological charge and
there are fermionic zero modes in the background of these solutions.
Interpretation of Euclidean solutions discussed in the present Letter
depend in particular on number of negative modes
in spectrum of small perturbations about them.
We plan to address  this topic in further work.

Another interesting topic for further study is the investigation
of Euclidean generalisations of axially symmetric \cite{kk96}
and platonic \cite{kk06} solutions of YMD theory.

\section*{Acknowledgements}
G.L. is grateful to the members of the Department of Theoretical Physics
of Mons University for kind hospitality during his visit to Belgium, when
the main part of this work was done. He wishes to
acknowledge the financial support of the Belgian F.N.R.S.
which made this visit possible.
He also would like to thank Hermann Nicolai for kind hospitality
during his visit to the Albert-Einstein-Institute, Golm, Germany,
where this work has been completed.



\newpage

\newpage

\newpage

\newpage

\newpage

\newpage

\newpage

\newpage

\begin{thebibliography}{99}

\bibitem{bm88}
R.~Bartnik and J.~Mckinnon,
``Particle - Like Solutions Of The Einstein Yang-Mills Equations,''
Phys.\ Rev.\ Lett.\  {\bf 61} (1988) 141.

\bibitem{lm93}
G.~V.~Lavrelashvili and D.~Maison,
``Regular and black hole solutions of Einstein Yang-Mills Dilaton theory,''
Nucl.\ Phys.\  B {\bf 410} (1993) 407.

\bibitem{biz93b}
P.~Bizon,
``Saddle points of stringy action,''
Acta Phys.\ Polon.\ B {\bf 24} (1993) 1209
[arXiv:gr-qc/9304040].

\bibitem{dg93}
E.~E.~Donets and D.~V.~Galtsov,
``Stringy sphalerons and nonAbelian black holes,''
Phys.\ Lett.\ B {\bf 302} (1993) 411
[arXiv:hep-th/9212153].

\bibitem{lm92}
G.~Lavrelashvili and D.~Maison,
``Static spherically symmetric solutions of a Yang-Mills field coupled to a
dilaton,''
Phys.\ Lett.\ B {\bf 295} (1992) 67.

\bibitem{biz93a}
P.~Bizon,
``Saddle point solutions in Yang-Mills dilaton theory,''
Phys.\ Rev.\ D {\bf 47} (1993) 1656
[arXiv:hep-th/9209106].

\bibitem{maison05}
D.~Maison,
``Static, spherically symmetric solutions of Yang-Mills dilaton theory,''
Commun.\ Math.\ Phys.\  {\bf 258} (2005) 657
[arXiv:gr-qc/0405052].

\bibitem{br06}
Y.~Brihaye and E.~Radu,
``Euclidean solutions in Einstein-Yang-Mills-dilaton theory,''
Phys.\ Lett.\ B {\bf 636} (2006) 212
[arXiv:gr-qc/0602069].

\bibitem{bl06}
Y.~Brihaye and G.~Lavrelashvili,
``Euclidean solutions of Yang-Mills-dilaton theory,''
arXiv:hep-th/0612238.

\bibitem{eg90}
A.~A.~Ershov and D.~V.~Galtsov,
``Nonexistence Of Regular Monopoles And Dyons In The SU(2) Einstein
Yang-Mills Theory,''
Phys.\ Lett.\ A {\bf 150} (1990) 159.

\bibitem{bp92}
P.~Bizon and O.~T.~Popp,
``No hair theorem for spherical monopoles and dyons in SU(2) Einstein
Yang-Mills theory,''
Class.\ Quant.\ Grav.\  {\bf 9} (1992) 193.

\bibitem{fg96}
P.~Forgacs and J.~Gyurusi,
``Static spherically symmetric monopole solutions in the presence of a
dilaton field,''
Phys.\ Lett.\ B {\bf 366} (1996) 205
[arXiv:hep-th/9508114].

\bibitem{wit77}
E.~Witten,
``Some Exact Multipseudoparticle Solutions Of Classical Yang-Mills  Theory,''
Phys.\ Rev.\ Lett.\  {\bf 38} (1977) 121.

\bibitem{fm80}
P.~Forgacs and N.~S.~Manton,
``Space-Time Symmetries In Gauge Theories,''
Commun.\ Math.\ Phys.\  {\bf 72} (1980) 15.

\bibitem{fg98}
P.~Forgacs and J.~Gyurusi,
``Magnetic monopole solutions with a massive dilaton,''
Phys.\ Lett.\ B {\bf 441} (1998) 275
[arXiv:hep-th/9808010].

\bibitem{dm81}
D.~Maison,
``Uniqueness Of The Prasad-Sommerfield Monopole Solution,''
Nucl.\ Phys.\ B {\bf 182} (1981) 144.

\bibitem{bfm92}
P.~Breitenlohner, P.~Forgacs and D.~Maison,
``Gravitating monopole solutions,''
Nucl.\ Phys.\ B {\bf 383} (1992) 357.

\bibitem{bfm95}
P.~Breitenlohner, P.~Forgacs and D.~Maison,
``Gravitating monopole solutions. 2,''
Nucl.\ Phys.\ B {\bf 442} (1995) 126
[arXiv:gr-qc/9412039].

\bibitem{colsys}
U. Ascher, J. Christiansen and R. D. Russell,
``A collocation solver for mixed order system of boundary value problems",
Mathematics of Computation {\bf 33} (1979), 639;
``Collocation software for boundary-value ODEs",
ACM Transactions {\bf 7} (1981), 209.

\bibitem{gl94}
G.~Lavrelashvili,
``Fermions in the background of dilatonic sphalerons,''
Mod.\ Phys.\ Lett.\ A {\bf 9} (1994) 3731
[arXiv:hep-th/9410178].

\bibitem{vg91}
D.~V.~Galtsov and M.~S.~Volkov,
``Sphalerons in Einstein Yang-Mills theory,''
Phys.\ Lett.\ B {\bf 273} (1991) 255.

\bibitem{mw92}
I.~Moss and A.~Wray,
``Black Holes And Sphalerons,''
Phys.\ Rev.\ D {\bf 46} (1992) 1215.

\bibitem{gs94}
G.~W.~Gibbons and A.~R.~Steif,
``Yang-Mills cosmologies and collapsing gravitational sphalerons,''
Phys.\ Lett.\ B {\bf 320} (1994) 245
[arXiv:hep-th/9311098].

\bibitem{kk96}
B.~Kleihaus and J.~Kunz,
``Axially symmetric multisphalerons in Yang-Mills-dilaton theory,''
Phys.\ Lett.\ B {\bf 392} (1997) 135
[arXiv:hep-th/9609180].

\bibitem{kk06}
B.~Kleihaus, J.~Kunz and K.~Myklevoll,
``Particle-like platonic solutions in scalar gravity,''
Phys.\ Lett.\ B {\bf 638} (2006) 367
[arXiv:hep-th/0601124].

\end{thebibliography}
\end{document}